\documentclass[%
 aip,
 amsmath,amssymb,
 reprint%
]{revtex4-1}

\usepackage{graphicx}
\usepackage{dcolumn}
\usepackage{bm}
\usepackage{caption}    
\usepackage{subcaption}    
\usepackage{array}  
\usepackage[hidelinks,pagebackref,colorlinks=true,allcolors=blue]{hyperref}	

\usepackage{xcolor}   
\usepackage{soul}   
\usepackage[export]{adjustbox}  

\expandafter\def\expandafter\normalsize\expandafter{%
    \normalsize%
    \setlength\abovedisplayskip{4pt}%
    \setlength\belowdisplayskip{8pt}%
    \setlength\abovedisplayshortskip{-4pt}%
    \setlength\belowdisplayshortskip{2pt}%
}   
\expandafter\def\expandafter\small\expandafter{%
    \normalsize%
    \setlength\abovedisplayskip{-10pt}%
    \setlength\belowdisplayskip{5pt}%
    \setlength\abovedisplayshortskip{-10pt}%
    \setlength\belowdisplayshortskip{0pt}%
}   

\usepackage{relsize}    

\usepackage[utf8]{inputenc}
\usepackage[T1]{fontenc}
\usepackage{mathptmx}
\usepackage{etoolbox}
\usepackage{environ}         

\usepackage{titlesec}
\titlespacing*{\section} {0pt}{4ex}{2ex}    

\usepackage[outdir=./]{epstopdf}

\DeclareUnicodeCharacter{2060}{!!!!I-AM-HERE!!!!!}   
\DeclareUnicodeCharacter{03BD}{!!!!I-AM-HERE!!!!!}   
\DeclareUnicodeCharacter{03C3}{!!!!I-AM-HERE!!!!!}   
\DeclareUnicodeCharacter{03BC}{!!!!I-AM-HERE!!!!!}   
\DeclareUnicodeCharacter{03C3}{!!!!I-AM-HERE!!!!!}   
\DeclareUnicodeCharacter{03C3}{!!!!I-AM-HERE!!!!!}   

\makeatletter
\def\@email#1#2{%
 \endgroup
 \patchcmd{\titleblock@produce}
  {\frontmatter@RRAPformat}
  {\frontmatter@RRAPformat{\produce@RRAP{*#1\href{mailto:#2}{#2}}}\frontmatter@RRAPformat}
  {}{}
}%
\makeatother

\makeatletter
\newcommand{\@citeX}[3]{{%
  \@ifundefined{b@#2}{\mbox{\reset@font\bfseries ?}%
    \G@refundefinedtrue
    \@latex@warning
      {Citation `#2' on page \thepage \space undefined}}%
  {#1\csname b@#2\endcsname#3}}}
\newcommand{\citeA}[1]{\@citeX{[Ref.~}{#1}{]}}
\newcommand{\citeB}[1]{\@citeX{Reference~}{#1}{}}
\newcommand{\citeC}[1]{\@citeX{}{#1}{}}
\makeatother

\newlength{\myl}
\let\origequation=\equation
\let\origendequation=\endequation

\RenewEnviron{equation}{
  \settowidth{\myl}{$\BODY$}                       
  \origequation
  \ifdimcomp{\the\linewidth}{>}{\the\myl}
  {\ensuremath{\BODY}}                             
  {\resizebox{\linewidth}{!}{\ensuremath{\BODY}}}  
  \origendequation
}

\begin{document}

\preprint{AIP/123-QED}

\title{Increased accuracy and signal-to-noise ratio through recent improvements in Infra-Red Video Bolometer fabrication and calibration}

\author{Fabio Federici}

\altaffiliation{E-mail address: fabio.federici@ukaea.uk}
\affiliation{ 
Oak Ridge National Laboratory, Oak Ridge, Tennessee 37831, USA
}%
\affiliation{ 
United Kingdom Atomic Energy Authority, Culham Centre for Fusion Energy, Culham Science Centre, Abingdon, Oxon, OX14 3DB, United Kingdom
}%

\author{Jack J. Lovell}%
\affiliation{ 
Oak Ridge National Laboratory, Oak Ridge, Tennessee 37831, USA
}%

\author{G. A. Wurden}%
\affiliation{ 
Los Alamos National Laboratory, Los Alamos, New Mexico 87545, USA
}%

\author{Byron J. Peterson}%
\affiliation{ 
National Institute for Fusion Science, 322-6 Oroshi-cho, Toki 509-5292, Japan
}%

\author{Kiyofumi Mukai}%
\affiliation{ 
National Institute for Fusion Science, 322-6 Oroshi-cho, Toki 509-5292, Japan
}%


\begin{abstract}

The Infra-Red Video Bolometer (IRVB) is a diagnostic equipped with an infra-red camera that measures the total radiated power in thousands of LOSs within a large field of view (FOV). Recently validated in MAST-U \cite{Federici2023}, it offers a high spatial resolution map of the radiated power in the divertor region, where large gradients are expected.

The IRVB’s sensing element comprises a thin layer of high Z absorbing material, typically Platinum, usually coated with Carbon to reduce reflections \cite{Peterson2008}.
It is here explored the possibility of using a relatively inert material like Titanium, that can be produced in layers up to 1$\mu$m compared to 2.5$\mu$m for Pt, and then coat it with Pt of the desired thickness (0.3$\mu$m per side here) and Carbon. This leads to a higher temperature signal (2 to 3 times), and better spatial resolution (about 4 times), resulting in higher accuracy in the measured power \cite{Peterson2008}. This assembly is also expected to improve foil uniformity, as the Pt layer is obtained via deposition rather than mechanical processes \cite{Mukai2016}.

Given its multi-material composition, measuring the thermal properties of the foil assembly is vital. Various methods using a calibrated laser as a heat source have been developed, analysing the temperature profile shape \cite{Sano2012,Mukai2018} or fitting the calculated laser power for different intensities and frequencies \cite{Federici2023}.
It is here presented a simpler approach, that relies on analysing the separate components of the foil heat equation for a single laser exposure in a given area. This can then be iterated over the entire foil to capture local deviations.

\end{abstract}

\maketitle


\section{Introduction}\label{introduction}

Resistive bolometers are used in tokamaks to measure the total radiated power integrated along one line of sight (LOS). Using many LOSs the emissivity profile across the entire plasma can be reconstructed.\cite{Ingesson2000} This is accurate in regions with low emissivity spatial gradient and whether high time resolution is required. If high spatial resolution is required a large number of intersecting LOSs are needed\cite{Meineri2023,Craciunescu2023}, increasing complexity and cost. A simpler way to achieve high spatial resolution is the use of an Infra-Red Video Bolometer (IRVB), that with a single diagnostic can cover a large field of view.\cite{Federici2023}
The IRVB is composed of a thin foil of absorber material enclosed in a tube that leaves it exposed to the plasma only through a small aperture. The foil heats up depending on the plasma emissivity and its temperature is measured via an infrared camera. Potentially every pixel of the IR camera represent a different LOS into the plasma, easily achieving thousands of LOSs with a single device.

The IRVB was already implemented in a series of devices and
was traditionally used to monitor core radiation. A significant component of this radiation is Bremsstrahlung, whose wavelength distribution depends strongly on the electron temperature $T_e$ (see fig\ref{fig:spectra}). Being the core the hottest portion of the plasma (order of keV), significant emission is present in the soft x-ray region. To absorb this emission the absorber has to be thick, worsening the thermal performance of the foil. In the x-point and divertor regions of the plasma, instead, temperatures are much lower ($<$100eV), with a Bremsstrahlung spectral distribution shifted to longer wavelengths and a much stronger influence of line radiation. The emission lines of elements and impurities desirable in a plasma (D,C,N, not W) are also at a low energy, therefore the can be much thinner when absorbing radiation from x-point and divertor regions.

In this paper the design for an IRVB absorber dedicated to the MAST-U core and divertor region will be shown, including an innovative use of coating technology. This new multi material approach makes inferring global thermal properties even more important, so the calibration technique is further refined compared to existing approaches.\cite{Federici2023,Sano2012}



\section{Foil thermal characteristics}\label{Thermal characteristics}

The IRVB foil is observed with an infrared camera, and its temperature is calculated with eq.\ref{eq:BBphotons1}

\small
\begin{equation}
\label{eq:BBphotons1}
\begin{aligned}
{\Phi} (T) =& i \int_{ {\lambda}_1 }^{ {\lambda}_2 } {\frac{2 \pi c } { {\lambda}^4 } \frac {1} { e^{\frac {hc} {\lambda k T}} -1} {d \lambda} } \:,\: 
T = {\alpha}_r ({\Phi})\\
C =& a_1 \cdot {\Phi} (T) + a_2 \:,\: a_1 = \hat{a_1} / \varepsilon_{cal} \\
T =& {\alpha}_r \left (\frac {C - C_0} {a_1} \frac {\varepsilon_{cal}} {\varepsilon} + {\Phi} (T_0) \right )
\end{aligned}
\end{equation}
\normalsize

where $i$ is the integration time, $\lambda$ the wavelength, $\lambda_1-\lambda_2$ the wavelength range allowed by the optics and ${\alpha}_r$ an interpolator used instead of the full integral. $\varepsilon$ and $\varepsilon_{cal}$ are the foil and calibration source emissivities respectively and $T$ the surface temperature. $a_1$ and $a_2$ are the proportional and additive calibration coefficients calculated as if $\varepsilon_{cal}$=1. The emissivities are explicitly shown in the temperature equation to showcase their direct impact on the signal level and calibration.

The absorber foil is so thin (order of microns) that the temperature across it equilibrates in hundreds to fractions of microseconds depending on foil design\footnote{The temperature equalisation across a slab via conduction in time can be approximated with a solution of the type \[ T= \mathlarger{\Sigma}_{n}^{} {c_n sin \left(2 \pi n \frac{x}{t_f} \right) exp \left[ -t/ \left( \left(\frac{t_f}{2 \pi n } \right)^2 \frac{\rho c_p}{k} \right) \right]} \] with $t_f$ thickness, $\rho$ density, $c_p$ heat capacity and $k$ heat conductivity. The slower component of the sum has time constant $\left(\frac{t_f}{2 \pi } \right)^2 \frac{\rho c_p}{k} $, and the end of the transition is conventionally assumed at 5 times the time constant.}, therefore heat transport across the foil reduces to 2D. This is formulated as per eq.\ref{eq:heat2d}.\cite{Federici2023}

\small
\begin{equation}
\label{eq:heat2d}
\begin{aligned}
P_{foil}=& P_{\frac {\partial T} {\partial t}}+P_{\Delta T}+P_{BB} = \varepsilon P_{plasma}\\
P_{\frac {\partial T} {\partial t}}=& \dfrac{k \: t_f}{\kappa} \dfrac{dT}{dt} \\
 P_{\Delta T} =& -k \: t_f \:  \left( \dfrac{\partial^2 T}{\partial x^2} + \dfrac{\partial^2 T}{\partial y^2} \right) \\
 P_{BB} =& 2 \: \varepsilon \: \sigma_{SB} \: (T^4 - T_0^4)
\end{aligned}
\end{equation}
\normalsize

where $k$ is thermal conductivity, $t_f$ thickness, $\kappa$ thermal diffusivity and $\sigma_{SB}$ Stefan-Boltzmann constant. The temperature can be calculated for every pixel of the IR camera, and the derivatives locally approximated with finite differences. Importantly $P_{foil}$ is the power absorbed by the foil, that differs from the power radiated by the plasma by the reflected and transmitted photons. The latter is most significant for higher energy photons, and is usually minimised by selecting a thick enough foil. The former is most significant for lower energy photons, especially for metallic absorbers, and it is lowered by applying a carbon coating. Reflectivity was measured directly with resistive bolometers\cite{Huber2007} and it was shown to be reduced from about 90 to 20$\%$ in the visible and IR region of the spectra\cite{VanEden2018}. For IRVBs it was shown that the application of a carbon coating increases the temperature reached by the foil for the same applied laser power\cite{Mukai2018}. 
For simplicity here foil absorption and emissivity are assumed as equal, yielding the first line in eq.\ref{eq:heat2d}. This is more valid for cold plasmas as the fraction of energy delivered by lower energy photons, closer to the IR wavelength range, is more important.

For relatively short pulses (like in MASTU and TCV) the main contributor to the foil absorbed power is the temporal variation. To maximise signal available to the camera (temperature) the ${k \: t_f}/{\kappa}$ coefficient should be as small as possible.
In order to achieve a high spatial resolution on the foil
 thermal diffusion should be inhibited, therefore ${k \: t_f}$ minimised.
On top of this it is clear from both eq.\ref{eq:BBphotons1} and \ref{eq:heat2d} that the foil emissivity should be as high as possible, to capture most of the available radiation from the plasma and to have the high camera counts for given temperature.
The expected signal level can be estimated with an analytical solution to the 2D heat transfer equation for the case of a thin layer of material exposed to a Gaussian laser. The peak temperature is given by \cite{Bauerle2011}

\small
\begin{equation}
\label{eq:laser on foil}
\begin{aligned}
\Delta T =& \frac {P_{foil} \omega^2} {4 k t_f} \left[  E_i \left( -\xi(1+4t^*) \right) - E_i \left( -\xi \right) \right] e^\xi \\
\Delta C \approx& {a_1} \frac {\varepsilon} {\varepsilon_{cal}} {{\alpha}_r}^{-1} \left( \frac {\varepsilon P_{laser} \omega^2} {4 k t_f}\left[  E_i \left( -\xi(1+4t^*) \right) - E_i \left( -\xi \right) \right] e^\xi \right) \\
t^* =& \frac{\kappa}{\omega}t , \xi = \frac{\eta^*}{2 {t_f}^*} , \eta^* = \frac{\eta \omega}{k} , {t_f}^* = \frac{t_f}{\omega} \\
\eta=&\varepsilon\sigma_{SB}\frac{T^4-{T_0}^4}{T-T_0}\approx\varepsilon\sigma_{SB}4{T_0}^3
\end{aligned}
\end{equation}
\normalsize

where $2\omega$ is the $1/e^2$ width of the laser, $P_{foil}$ the power density absorbed by the foil, $E_i$ is the exponential integral function and $t$ is the time from the beginning of the exposure to the laser. $\Delta C$ is the increase of camera counts, estimated using the inverse of the correlation used in eq.\ref{eq:BBphotons1} (${{\alpha}_r}^{-1}$), using the same assumption on the difference between absorbed and input laser power from eq.\ref{eq:heat2d}.
\footnote{The double dependency on $\varepsilon$ appears similarly in the power calculated with eq.\ref{eq:heat2d}. Considering ${{\alpha}_r}$ linear, the emissivity in the determination of the temperature aligns with the one in the calculation of the power from the plasma in all terms except for the black body component, that has anyway a minor importance for pulsed devices.} $\eta$ represent the power losses on the surface. Given the foil is in vacuum these reduce to black body radiation, so this becomes $\eta \delta T$. $\eta$ is not constant but approximated as shown the error is less than 6\% for a temperature increase of 10K above room temperature.

Eq.\ref{eq:laser on foil} returns similar results as more sophisticated model in \cite{Mukai2021} and can be used to more readily estimate the signal level. The temperature signal can be directly compared to the noise equivalent temperature difference (NETD, typically $\sim$0.03K) of the infrared camera, a metric additional to the estimation of the noise equivalent power density (NEPD, measured 12$W/m^2$ compared to 10 calculated) given by eq.\ref{eq:nepd}\cite{Pandya2014b}.

\small
\begin{equation}
\label{eq:nepd}
\begin{aligned}
NEPD = \frac{ \sqrt{10} k t_f \sigma_{IR} }{\sqrt{f_{IR} N_{IR}}} \sqrt{ \frac{N_{bol}^3 {f_{bol}}}{A_f^2} + \frac{N_{bol} {f_{bol}^3}}{5 \kappa^2}}
\end{aligned}
\end{equation}
\normalsize

The power density $P_{laser}$ and spatial scale $\omega$ of the radiation on the foil can be estimated from existing plasma simulations with the ray tracing code Cherab\cite{Carr2024}
as done in \cite{Federici2023,Federici2023b}. The exposure time $t$ can be assumed as the time step given by the camera framerate after binning.

A low thickness always leads to higher signal. At a high framerate diffusion does not have enough time to act, so the most relevant factor is the local thermal inertia of the foil, while at low framerate the conductivity dominates. For MASTU conditions, among the exhaustive list of materials in \cite{Mukai2021}, the ones that return the highest signal with a fixed 1$\mu$m thickness are Titanium, Zirconium, Niobium, Cadmium, Indium, Tin, Hafnium and Tantalum. Of these, a high steady state temperature (useful if the estimated signal is too low and a larger time binning is necessary) is given by Ti, Zr and Hf.

A limiting factor in selecting the most appropriate material is the minimum thickness of foil that can be produced and be self supporting. \cite{Mukai2021} provides a list of achievable thicknesses (with this the best materials become Ti, Zr and Pd) but this is contingent to the evolution of manufacturing technology. The foils are usually produced by rolling, a mechanical process practical only up to a given foil width before it tears. Another limitation is that the imperfections in the rollers reflect to non uniformity of the foil, that become more and more relevant as the thickness decreases\cite{Federici2023} to the point where the rollers can touch and create a hole (pinhole)
. The practical thickness limit depends on the material but is usually $\geq$1$\mu$m. 

As mentioned before most foils are coated with carbon to reduce the reflectivity to low energy photons, but this effects the thermal performance. The thickness of a layer of carbon deposited via spay is around 5$\mu$m\cite{Pandya2014} while if applied via vapour deposition it was shown that only $\sim$200nm are needed.\cite{Mukai2018} The deposition technique has an impact also on the foil uniformity, as it is difficult to create homogeneous layers by spraying, and the thermal adhesion of the sprayed particles is uncertain, potentially contributing to the non uniformity observed in \cite{Federici2023}.

\section{Foil photon absorption}\label{Photon absorption}

All discussed above regards the foil thermal properties, but the foil function is first to absorb the photons coming from the plasma. The photon absorption depends on the material and is heavily wavelength dependent. Low energy photons, albeit susceptible to reflection, are easily absorbed, while this is harder for higher energy ones. The thickness requirement will therefore be heavily dependent on the spectra of the radiation to be observed. The main sources of radiation in tokamaks are Bremmstrahlung (free-free), recombination (free-bound and bound-bound) and line radiation (bound-bound).

SOLPS simulations are used as reference cases for typical plasma conditions. The SOLPS domain is limited to the edge, so the solution has to be extended to the core manually or with other codes.\cite{Peterson2022} This data can be fed to Cherab, that calculates the line integral relying on the OpenADAS collisional radiative database.
\footnote{Currently Cherab cannot model free-bound radiation from recombination wavelength resolved. For the examined case, though, the wavelength integrated power due to free-bound recombination radiation is about 25\% of the Bremsstrahlung component and, in virtue of the stronger negative temperature dependency ($\alpha (\frac{E_r}{k_B T_e})^{\frac{3}{2}}$ rather than $(\frac{E_r}{k_B T_e})^{\frac{1}{2}}$), it is most efficiently generated in colder plasma regions, therefore yielding lower energy photons. Not including the recombination free-bound radiation in the spectra is therefore likely conservative.}
For MASTU a series of Super-X H-mode SOLPS simulations with a fuelling and seeding scan from \cite{Myatra2023} were used
. Plasma density and temperature have been linearly extended to arbitrary $10^{20} \#/m^3$ and 2.5keV respectively (these are conservative assumptions compared to current measurements), the impurity concentration was extended from the core edge using the Collisional Ionisation Equilibrium\cite{Kaastra2008}, maintaining the total impurity over plasma density constant.\footnote{This could introduce a non uniform $Z_{eff}$, but given the high temperature inside the SOLPS core edge the impurities are practically fully ionised, so $Z_{eff}$ is effectively constant.} To examine the most representative LOSs, one directed from the pinhole to the magnetic axis and one directed to the X-point have been selected.
Below the X-point the temperature decreases significantly, implying lower energetic photons and conditions even more favourable to their absorption.

\begin{figure}
	\centering
	\includegraphics[trim={5 10 10 45},clip,width=0.85\linewidth]{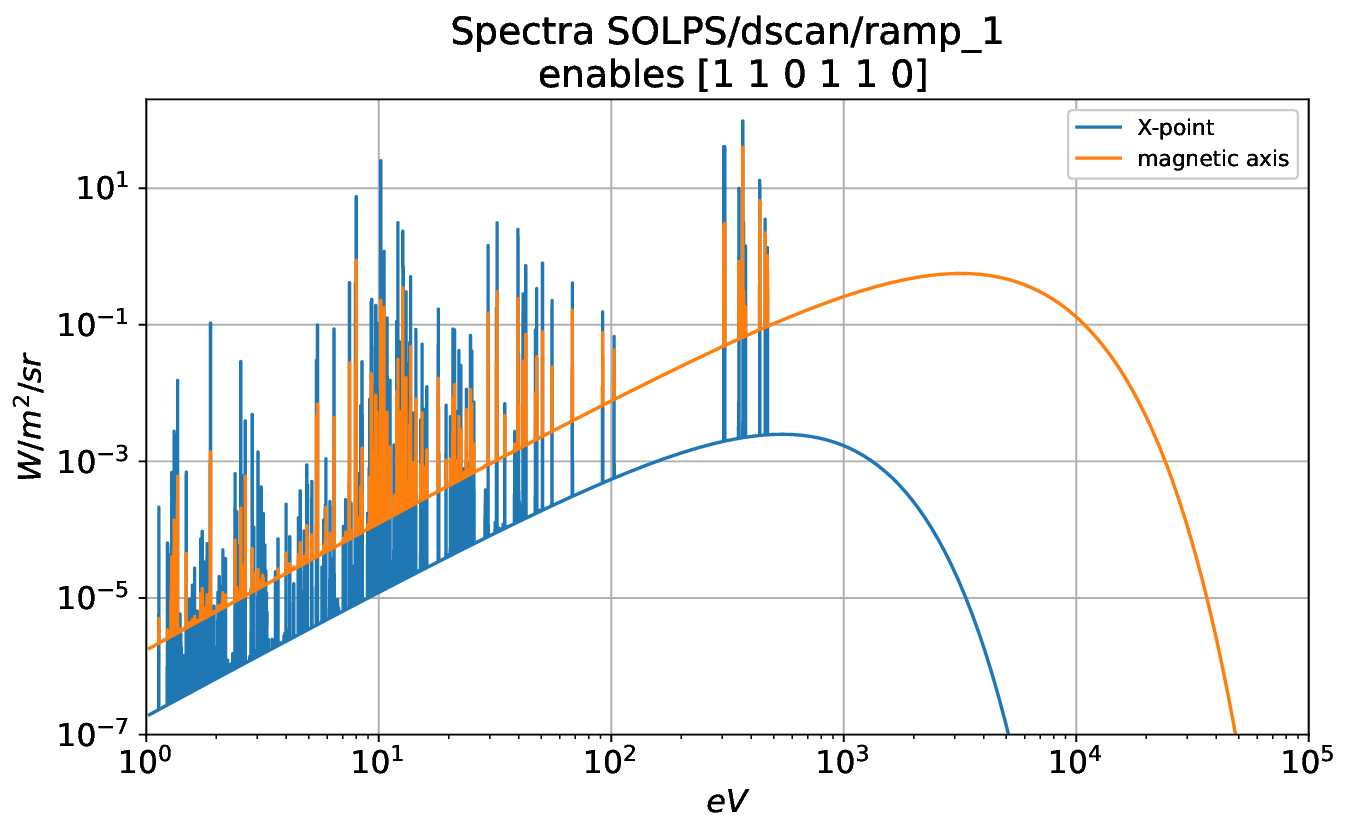}
         \vspace{-3mm}
	\caption{Line integrated spectra for a low fuelling high power SOLPS simulation, calculated with Cherab, for LOSs from the IRVB pinhole and directed as per legend. The Cherab bin size in across the wavelength range was set constant to 0.001 nm, therefore inversely proportional to eV.}
	\label{fig:spectra}
\end{figure}

Among the available simulations, one with low fuelling and without Nitrogen has the highest fraction of radiated power delivered by high energy photons ($>$5keV), because of the lack of N lines and high edge $T_e$. The relative spectra is shown in fig.\ref{fig:spectra}. It can be seen the large contribution of Bremmstrahlung compared to the emission lines, but the energy from high energy photons is still $<$6\% of the total. The sole component from line radiation is between 16\% and 40\% for the magnetic axis LOS, while around 98\% for the X-point one.

\begin{figure}
	\centering
	\includegraphics[trim={5 10 10 45},clip,width=0.8\linewidth]{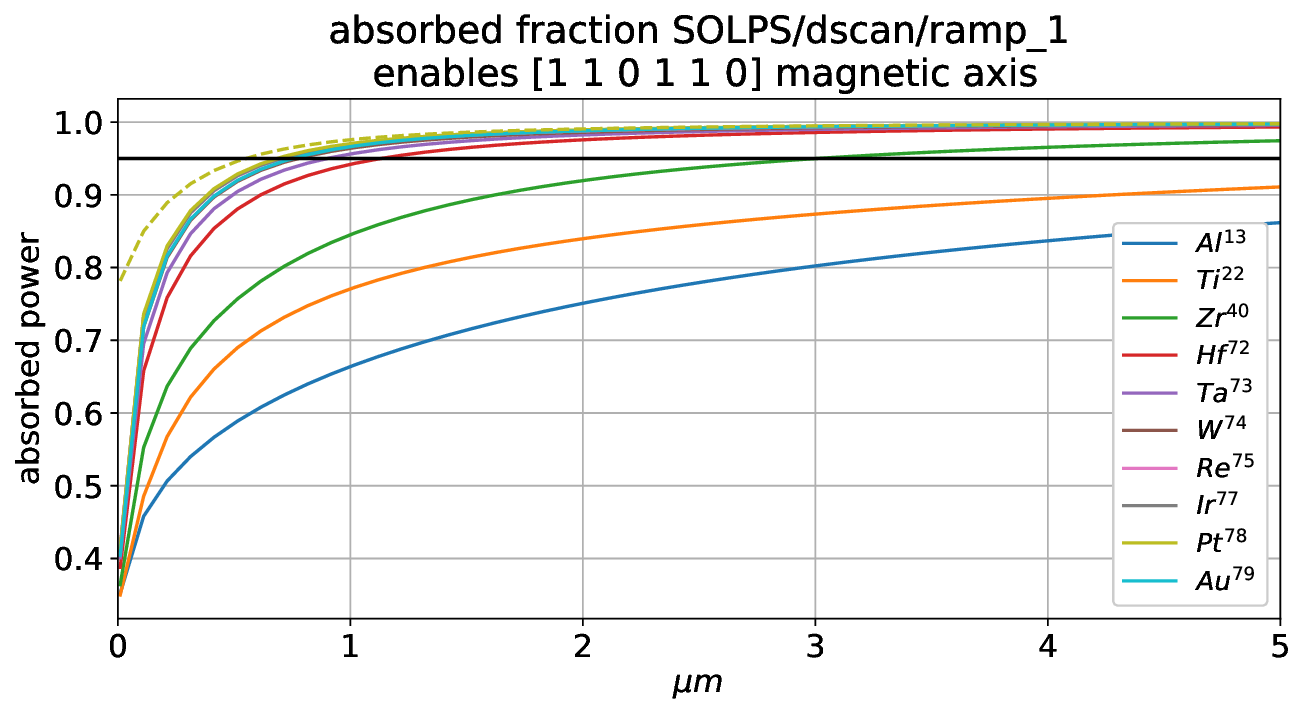}
        \vspace{-3mm}
	\caption{Fraction of the energy absorbed by the foil with thickness for different constituent elements. These include a 250nm carbon layer per side. Dashed is the reduction in Pt thickness given by considering the 1$\mu$m Ti substrate. The black line indicates the 95\% selection criteria as per \cite{Peterson2022}.}
	\label{fig:absorption}
\end{figure}

The wavelength dependent photon attenuation tables per element are available at \cite{Chantler2005}.
The fraction of energy absorbed as function of material thickness is shown in fig.\ref{fig:absorption}. As already highlighted in \cite{Mukai2021}, to reach a high absorption with high Z elements the thickness required is lower than the minimum producible one. With low Z materials instead the thickness would be so high to impact the foil thermal performance. 

It is here proposed to combine low and high Z materials in the same foil.
The heavy elements can be vapour deposited on a thin layer made of a light one. The former acts as absorber, and only as little as required can be deposited without incurring into the risks of non uniformity.\cite{Mukai2016,Federici2023} The latter acts as mechanical substrate. Given the low thickness, this could be effected by non uniformity defects
. The thermal performance of the thinnest producible low Z foil and the deposited high Z layer is comparable. The high uniformity of the latter will help alleviate the non uniformity of the former.

Other factors to consider are the compatibility with respect to magnetic fields, adhesion between layers, stability in air, vacuum and to neutron irradiation. All considered the best material as absorber is Platinum, that is already commonly used for IRVBs and compatible with Carbon (and Titanium, see next paragraph). It must be noted, though, that this is only valid for high time resolution, as for higher time binning Hafnium can become advantageous. To the author knowledge Hafnium was not tested before in a tokamak environment, so its compatibility is uncertain.

The most promising low Z substrate is Titanium, producible mechanically up to 1$\mu$m, while the others only in higher thickness. The Platinum thickness required to achieve the threshold absorption including the Titanium layer reduces from 0.69 to 0.55$\mu$m. It must be noted though that this choice is contingent 
on what foil is commercially available. One promising manufacturing method for the substrate relies on depositing material over a soluble sacrificial layer, that is then dissolved. The resulting foil can be thinner than what achievable mechanically while maintaining the uniformity characteristic of vapour deposition. At the time of writing the only suitable foil found was Aluminium of a thickness of 0.4$\mu$m by American Elements. This, for a higher cost, outperforms 1$\mu$m Ti for a large range of high framerate applications in virtue of the lower density. Titanium is not compatible with vapour deposited carbon, so the Platinum layer is also applied on both sides of the Titanium substrate.

It must also be noted that the minimum thickness for 95\% absorption is heavily dependent on the particular LOS examined. Here, to allow for a possible integration with the existing resistive bolometer system\cite{Lovell2023}, LOSs through the core have been considered. With a lower peak $T_e$ the necessary thickness decreases. The Pt thickness required on top of the Ti layer becomes with 0.75$\mu$m with peak $T_e$=3keV (close to what can be achieved with conventional production methods) while 0.39$\mu$m with peak $T_e$=2keV. Restricting the FOV at and below the X-point causes the low Z substrate alone to be sufficient. Ti is also less resilient to neutron irradiation than Pt, so it's use in devices operating in D+T has to be carefully evaluated.

\section{Foil calibration}\label{Calibration}

Once the foil is constructed its thermal properties have to be measured. Essential are: a calibrated IR camera, a vacuum system to contain the foil and a localised heat source (laser).

The IR camera calibration is usually performed with a Back Body (BB) source smaller than the camera FOV. $a_1$ is then assumed constant across the FOV and a spatial distribution for $a_2$ is found with a non uniformity correction (NUC) plate.
A more recent attempt to capture the camera non uniformities involved heating and cooling a NUC plate while monitoring its temperature.\cite{Federici2023} This attempt could capture variations in $a_1$, but introduces an additional uncertainty as the NUC plate, cooled by air convection, does not have a uniform temperature (estimated up to 0.25K difference).
To obtain more stable and repeatable results the more rigorous system seems to be the use of a BB source with an active area as large as the camera FOV, like what supplied by HGH Infrared Systems.

To calibrate the foil various methods have been developed. In a family of attempts the steady state spatial temperature distribution around a laser spot is compared with a simulation of the temperature using a finite element method finding the best foil parameters ($t_f$, $\epsilon$). $\kappa$ is then found by fitting the foil cooling after the laser heating is removed. The laser is then moved across the foil to scan its entire surface.\cite{Sano2012,Mukai2016,Itomi2014}
This procedure relies numerical simulations, and did not include the emissivity dependency of the foil temperature in eq.\ref{eq:BBphotons1} and in the first line of eq.\ref{eq:heat2d}, assuming them unitary.

A later attempt consisted in performing different laser exposures in the same location varying the laser power and focus. The power on the foil is then calculated with eq.\ref{eq:BBphotons1} and \ref{eq:heat2d} within a set area, finding the parameters that return the best match across all conditions.\cite{Federici2023} This returned reasonable results, but later examination showed that changing the integration radius would significantly change $P_{foil}$, clearly not possible. This also did not include the mentioned emissivity dependency.

It is here proposed a different approach, that relies on a single laser exposure, and does not rely on a complex data interpretation. \footnote{Another possible approach, not explored here, is to fit the foil peak temperature with the expression in eq.\ref{eq:laser on foil}. Testing this idea showed that the solution is very dependent on the extremes of the fit, and the approximation on $\eta$ would fail at high temperatures. An attempt could be made with a very slow signal and low laser power.} It will be applied to the data regarding the IRVB foil currently installed in MASTU for the experimental campaigns 1 to 4. 

When a laser is shone on the foil the temperature increase is very localised. From there, heat diffusion will transport the heat to the rest of the foil. This process requires time to occur so, at the very beginning of the laser exposure, no heat is transferred outside a small area around the laser spot. 
Within this time and area the integral of  $P_{\Delta T}$ in eq.\ref{eq:heat2d} cancels out, so only $P_{\partial T / \partial t}$ and $P_{BB}$ remain. 


For a given $\epsilon$ and $T_0$ the temperature and $P_{BB}$ can be determined, and the value of the coefficient $ k \: t_f / \kappa$ can be directly calculated with eq.\ref{eq:coeff_1}

\begin{figure}
	\centering
	\includegraphics[trim={5 5 5 5},clip,width=0.9\linewidth]{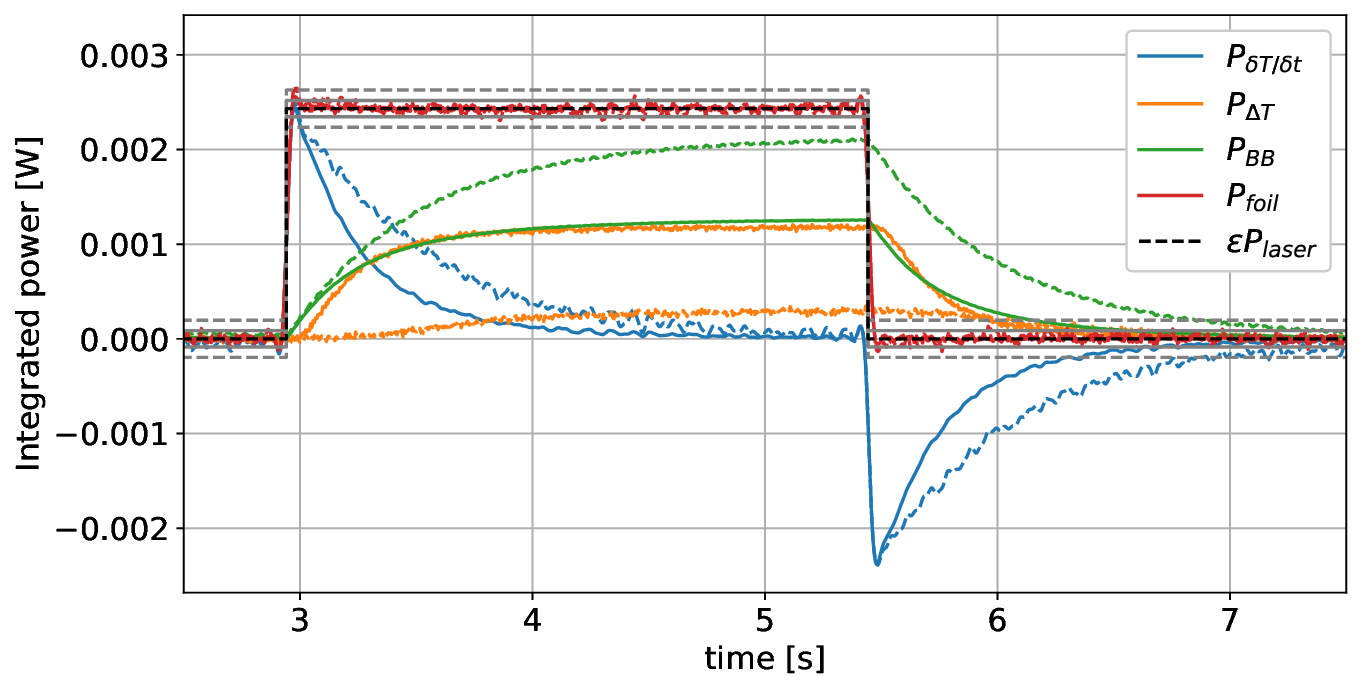}
        \vspace{-3mm}
	\caption{In dashed black the nominal laser input power profile absorbed by the foil, while the others are the calibrated components and total of $P_{foil}$ integrated within 3.4mm (solid lines) and 7.5mm (dashed) of the laser spot. The time derivative signals have been smoothed to limit the effect of the known constant oscillation in the count signal described in \cite{Federici2023}. The gray lines show the noise level calculated with eq.\ref{eq:nepd} around the input power level.}
	\label{fig:not fitted power}
\end{figure}

\small
\begin{equation}
\label{eq:coeff_1}
\begin{aligned}
 \frac{k \: t_f }{\kappa} =  \frac{ \varepsilon P_{laser} - \int_{A} P_{BB} } {\int_{A} \frac{dT}{dt} }
\end{aligned}
\end{equation}
\normalsize

evaluated at the peak of $P_{\partial T / \partial t}$ ($A$ indicates the area of interest). fig.\ref{fig:not fitted power} shows an example of the individual contributions to the known $P_{laser}$, and how the $P_{\Delta T}$ component lags behind the true laser signal.\footnote{$P_{BB}$ changes are the best metric to reconstruct the input laser wave shape.} 

The coefficient of the diffusion component can be found at the end of the laser exposure, when the temperature reached steady state, with eq.\ref{eq:coeff_2}.

\small
\begin{equation}
\label{eq:coeff_2}
\begin{aligned}
k \: t_f  =  \frac{ \varepsilon P_{laser} - \int_{A} P_{BB}  - \int_{A} P_{\frac {\partial T} {\partial t}} } {\int_{A} \left( \frac{\partial^2 T}{\partial x^2} + \frac{\partial^2 T}{\partial y^2} \right) }
\end{aligned}
\end{equation}
\normalsize

\begin{figure*}[!ht]
    \captionsetup{labelfont={color=white}}
     \centering
     \begin{subfigure}{0.33\textwidth}
        \centering
        \vspace*{-0mm}
	  \includegraphics[trim={10 5 30 40},clip,width=0.8\linewidth]{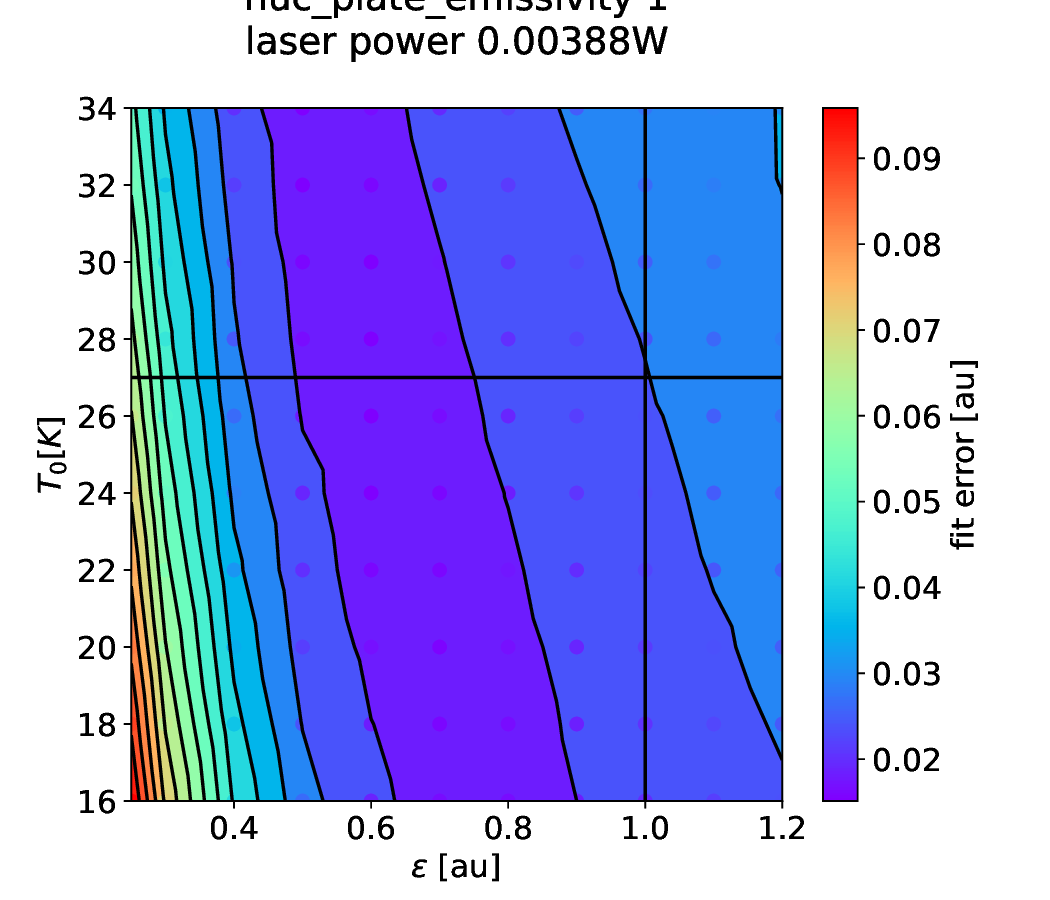}
         \vspace*{-10mm}
         {\color{white}\caption{\phantom{weww}}\label{fig:fita}}
     \end{subfigure}
     \hfill
     \begin{subfigure}{0.295\textwidth}
         \centering
         \vspace*{-0mm}
	   \includegraphics[trim={60 5 30 40},clip,width=0.8\linewidth]{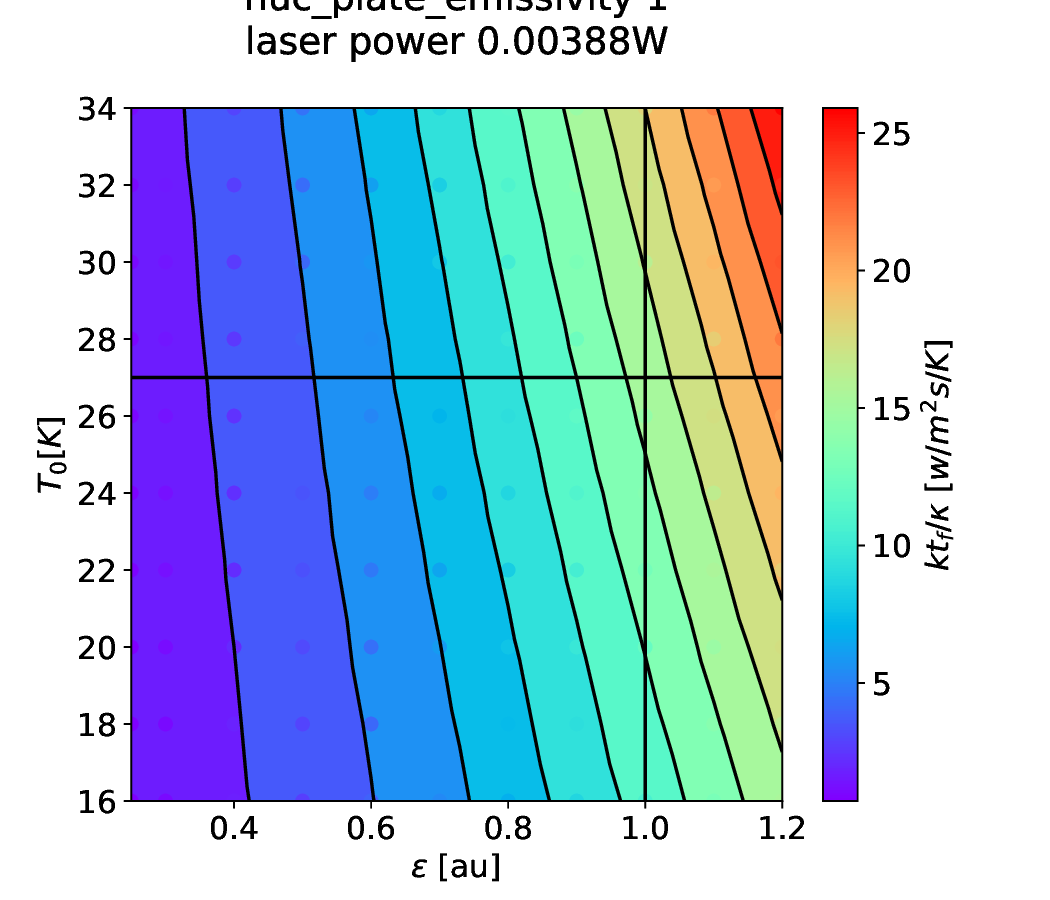}
         \vspace*{-10mm}
         {\color{white}\caption{\phantom{wewwwww}}\label{fig:fitb}}
     \end{subfigure}
     \hfill
     \begin{subfigure}{0.31\textwidth}
         \centering
         \vspace*{-0mm}
	   \includegraphics[trim={60 5 5 40},clip,width=0.8\linewidth]{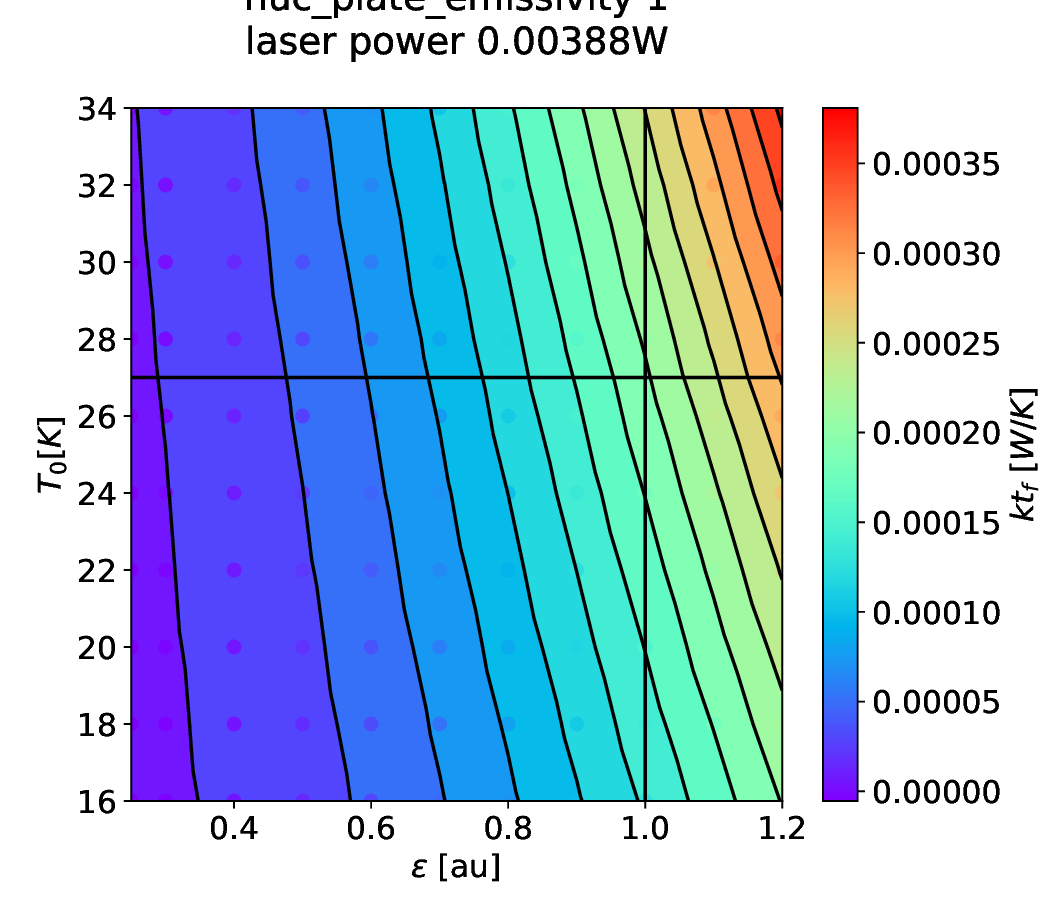}
         \vspace*{-10mm}
         {\color{white}\caption{\phantom{wewwwww}}\label{fig:fitc}}
     \end{subfigure}
        \captionsetup{labelfont={color=black}}
        \vspace*{+6mm}
        \caption{Fit of $\varepsilon P_{laser}$ with $P_{foil}$ for the smaller area around the laser spot, calculated with $\varepsilon_{cal}=1$. (\subref{fig:fita})Sum of the square difference error, (\subref{fig:fita})time derivative and (\subref{fig:fita})diffusion coefficients.}
        \label{fig:fit}
\end{figure*}

This completely defines the foil properties for a set pair of $\epsilon$ and $T_0$.\footnote{The emissivity of the calibration source $\varepsilon_{cal}$ is a hidden parameter within eq.\ref{eq:BBphotons1}, and has to be known beforehand. This analysis cannot be used to find $\varepsilon_{cal}$, that influences mostly the final values of $\varepsilon$ and $k$ while not $\kappa$.} To find the correct values, the difference between the known $\varepsilon P_{laser}$ and $P_{foil}$ can be calculated throughout the laser pulse. This operation works also if using only one area as reference. The result is a plot of the error as in fig.\ref{fig:fita}.

The variation with $T_0$ is weak so, if not known, the values of $\varepsilon$, $\frac{k \: t_f }{\kappa}$, $k \: t_f$ can be found by averaging the properties associated with the region of lower error, the standard deviation representing the uncertainty (10-15\%). Using Pt nominal conductivity, these are as follows: $\varepsilon=0.63$, $t_f=0.71 \mu m$, $\kappa = 9.5E^{-6} m^2/s$. These are different from previous estimations\cite{Federici2023,Reinke2018a,Itomi2014} but the thickness matches very closely what was directly measured by weighing a known area of the same uncoated foil batch (0.72$\mu m$). It has to be reminded that this result is heavily dependent on the camera calibration, that will be improved with the new black body source. This calibration was tested for other laser power, frequency and integration area than the one it was determined from, and the result match the expectation.

\section{Summary}\label{Summary}

In this paper the design criteria for a IRVB foil have been discussed, with particular emphasis to how the signal levels can be improved. If the IRVB is designed to aim at relatively cold regions of the plasma, or the peak temperature is sufficiently low, the fraction of energy from high energy photons is low and the thickness of the foil can be reduced. A model for the signal level was proposed based on an analytical solution of the heat diffusion equation. This highlights that different materials respond differently to transients, so the choice must be operated based both on the radiation source but also the IR camera of choice. To work around the current limitations in foil production technology, vapour deposition of the absorber on top of a material that provides mechanical support leads to better thermal performance. The likely best substrate material is Titanium, or depending on production availability Aluminium and Zirconium. Platinum or Hafnium are both good absorbers, even if Hafnium is less established.

To calculate the foil thermal properties, a simple and direct approach is here proposed, relying on a single laser exposure at high power and very low frequency. The result seems closer to a test of the weight of the same batch of uncoated foil, and it works for different laser condition and integration area.

\begin{acknowledgments}

This work is supported by US Department of Energy, Office of Fusion Energy Sciences under the Spherical Tokamak program, contract DE-AC05-00OR22725.

The authors wish to also acknowledge Felix Reimold and Gabriele Paresotti for the useful conversation on the topic of the IRVB.

\end{acknowledgments}

\section{References}
\bibliographystyle{IEEEtran}
\bibliography{references_short}

\appendix

\end{document}